\documentclass[preprintnumbers]{revtex4}

\usepackage{amsfonts}
\usepackage{amsmath}
\usepackage{hyperref}
\usepackage{amssymb}
\usepackage[english]{babel}
\usepackage{graphicx}
\usepackage{epsfig}
\usepackage{bm}
\usepackage{longtable}
\usepackage{verbatim}
\usepackage{longtable}

\newcommand{\mathsym}[1]{{}}

\input{tcilatex}

\begin{document}

\title{Rare top decay and CP violation in THDM}

\author{R. Gait\'an${}^{a}$ }
\email{rgaitan@unam.mx}     
\author{R. Martinez${}^{b}$}
\email{remartinezm@unal.edu.co}     
\author{ J.H. Montes de Oca Y.${}^{a}$}
\email{josehalim@comunidad.unam.mx}      
\author{ S. Rodriguez Romo${}^{c}$}
\email{suemi@unam.mx}

 \affiliation{ Departamento de Fisica, FES-Cuautitl\'an, Universidad Nacional Auton\'oma de M\'exico, \\
C.P. 54770, Estado de M\'exico, M\'exico${}^{a}$  \\
 Departamento de Fisica, Universidad Nacional de Colombia, Bogot\'a D.C., Colombia${}^{b}$  \\
          Departamento de Quimica, FES-Cuautitl\'an, Universidad Nacional Auton\'oma de M\'exico, C.P. 54770, Estado de M\'exico, M\'exico${}^{c}$ }

\date{today}

\begin{abstract}
We discuss the formalism of two Higgs doublet model type III with CP violation from CP-even CP-odd mixing in the neutral Higgs bosons. The flavor changing interactions among neutral Higgs bosons and fermions are presented at tree level in this type of model. These assumptions allow the study rare top decays mediated by neutral Higgs bosons, particularly we are interested in $t\rightarrow c l^+l^-$. For this process we estimated upper bounds of the branching ratios $\textrm{Br}(t\rightarrow c \tau^+\tau^-)$ of the order of $10^{-9}\sim 10^{-7}$ for a neutral Higgs boson mass of 125 GeV and $\tan\beta=1$, 1.5, 2, 2.5. For the case of $t\rightarrow c \tau^+\tau^-$ the number of possible events is estimated from 1 to 10 events which could be observed in future experiments at LHC with a luminosity of 300 $\textrm{fb}^{-1}$ and 14 GeV for the energy of the center of mass. Also we estimate that the number of events for the process $t\rightarrow c l^+l^-$ in different scenarios is of order of $2500$.
\end{abstract}

\maketitle

%
%\pacs{12.60.Fr, 12.15.Mm, 14.80.Cp}%%%%%%%%%   CAMBIAR PACS
%
%
%%%%%%%%%%%%%%%%%%%%%%%%%%%%%%%%%%%%%%%%%%%%%%%%%%%%%%%%%%%%%%%%%%%% SECTION 1 %%%%%%%%%%%%%%%%%%%%%%%%%%%%%%%%%%%%%%%%%%%%%%%
\section{Introduction}
The last results from LHC have confirmed the observation of one scalar particle with mass on the electro-weak scale. The ATLAS \cite{atlas125gev} and CMS \cite{cms125gev} collaborations have been reported the observation of a new particle with mass of around to 125 GeV. The observation has an important significance of more than 5 standard deviations. Even with this research it is not yet possible for us to name this particle as the Standard Model Higgs boson. However if this result is confirmed by future analysis, it will be one of the greatest discoveries of mankind. On the other hand, the SM is often considered as an effective theory, valid up to an energy scale of $O(GeV)$, that eventually will be replaced by a more fundamental theory, which will explain, among other things, the physics behind electro-weak symmetry breaking and perhaps even the origin of flavor. Many examples of candidate theories, which range from supersymmetry \cite{susyrev1,susyrev2} to strongly interacting models \cite{ArkaniHamed:2001nc} as well as some extra dimensional scenarios \cite{Chang:2010et}, include a multi-scalar Higgs sector. In particular, models with two scalar doublets have been studied extensively \cite{BrancoReport}, as they include a rich structure with interesting phenomenology.

First versions of the two Higgs doublet model (2HDM) are known as 2HDM-I \cite{2hdmI1,2hdmI2} and 2HDM-II \cite{2hdmII}. These versions involve natural flavor conservation and CP conservation in the potential through the introduction of a discrete symmetry. A general version which is named as 2HDM-III allows the presence of flavor-changing scalar interactions (FCNSI) at a three level \cite{2hdmIII}. There are also some variants (known as top, lepton, neutrino), where one Higgs doublet couples predominantly to one type of fermion
\cite{BrancoReport}, while in other models it is even possible to identify a candidate for dark matter \cite{2hdm:darkmatter1,2hdm:darkmatter2}. The definition of all these models, depends on the Yukawa structure and symmetries of the Higgs sector, whose origin is still not known. The possible appearance of new sources of CP violation is another characteristic of these models \cite{2dhmCPV}.

Within 2HDM-I where only one Higgs doublet generates all gauge and fermion masses, while the second doublet only knows about this through mixing, and thus the Higgs phenomenology will share some similarities with the SM, although the SM Higgs couplings will now be shared among the neutral scalar spectrum. The presence of a charged Higgs boson is clearly the signal beyond the SM. Within 2HDM-II one also has natural flavor conservation \cite{Glashow:1976nt}, and its phenomenology will be similar to the 2HDM-I, although in this case the SM couplings are shared not only because of mixing, but also because of the Yukawa structure. The distinctive characteristic of 2HDM-III is the presence of FCNSI, which require a certain mechanism in order to suppress them, for instance one can imposes a certain texture for the Yukawa couplings \cite{Fritzsch:1977za}, which will then predict a pattern of FCNSI Higgs couplings \cite{2hdmIII}. Within all those models (2HDM I,II,III) \cite{Carcamo:2006dp}, the Higgs doublets couple, in principle, with all fermion families, with a strength proportional to the fermion masses, modulo other parameters.

With higher energy, as planned, the LHC will also become an amazing top factory, allowing to test the top properties, its couplings to SM channels and
rare decays \cite{roberto:top}. One of the interesting rare decays for the top is $t\rightarrow c l^+l^-$, which is a clear signal of new physics. In literature this type of top decay is often known as rare top decay and it could be mediated at three level by neutral gauge bosons in the context of physics beyond SM. For instance, models with additional gauge symmetries introduces an neutral gauge boson $Z^\prime$ which allows the rare top decay \cite{Zprime1,Zprime5,Zprime6}. The obtained results for branching ratios with flavor changing neutral currents are extremely suppressed due to the mass of additional gauge boson $Z^\prime$ which must be of the order of TeV. However, in the framework of the 2HDM-III these rare top decay are possible at three level through neutral Higgs bosons in the framework of general 2HDM with upper bounds of branching ratio $t\rightarrow c l^+l^-$ less suppressed.

In this work we discuss the flavor changing neutral Higgs interactions due to Yukawa couplings and a CP violation source from Higgs sector in the framework of 2HDM-III. Our analysis is devoted to the study of decay $t\rightarrow c l ^+ l^-$ at tree level with basic goal of identifying effects of new physics. The organization of the paper goes as follows: Section \ref{sec2} describes the CP violation source in Higgs sector. The flavor changing interaction between neutral Higgs bosons and fermions are introduced in section \ref{sec3}. Section \ref{sec4} contains the analysis of the branching ratio for rare top decay. Finally section 5 we present our conclusion and discussion.
%
%
%%%%%%%%%%%%%%%%%%%%%%%%%%%%%%%%%%%%%%%%%%%%%%%%%%%%%%%%%%%%%%%%%%%% SECTION 2 %%%%%%%%%%%%%%%%%%%%%%%%%%%%%%%%%%%%%%%%%%%%%%%
\section{Neutral Higgs bosons spectrum}
\label{sec2}
Let $\Phi_1$ and $\Phi_2$ denote two complex $SU(2)_L$ doublet scalar fields with hypercharge-one. The most general gauge invariant and renormalizable Higgs scalar
potential in a covariant form with respect to global $U(2)$
transformation is given by \cite{oneil}
\begin{equation}
V=Y_{a,\overline{b}}\Phi_{\overline{a}}^\dag\Phi_b+\frac{1}{2}Z_{a\overline{b}c\overline{d}}\left(\Phi_{\overline{a}}^\dag\Phi_b\right)
\left(\Phi_{\overline{c}}^\dag\Phi_d\right),
\end{equation}
where $\Phi_a=\left(\phi_a^+,\,\phi_a^0 \right)^T$ and $a,b,c,d$ are
labels with respect to two dimensional Higgs flavor space. The index
conventions means that replacing an unbarred index with a barred
index is equivalent to complex conjugation and barred-un\-ba\-rred indices
denote a sum. The most general $U(1)_{EM}$-conserving vacuum expectation values are
\begin{equation}
\langle \Phi_a \rangle= \frac{1}{\sqrt{2}}\left(
\begin{array}{c}
0 \\
v_a \\
\end{array}
\right),
\label{vev}
\end{equation}
where $\left(\begin{array}{cc}v_1, & v_2 \\ \end{array}\right) =\left(\begin{array}{cc} v\cos \beta,& v\sin\beta
\\ \end{array}\right)$ and $v=246$ GeV.

After spontaneous symmetry breaking, an orthogonal transformation $R$ is used to diagonalize the squared mass matrix for neutral Higgs fields. The mass-eigenstates of the neutral Higgs bosons are
\begin{equation}
h_{i}=\sum_{j=1}^{3}R_{ij}\eta _{j},
\label{h-Rn}
\end{equation}
where $R$ can be written down as:
\begin{equation}
R=\left(
\begin{array}{ccc}
c_{1}c_{2} & s_{1}c_{2} & s_{2} \\
-\left( c_{1}s_{2}s_{3}+s_{1}c_{3}\right)
& c_{1}c_{3}-s_{1}s_{2}s_{3} & c_{2}s_{3} \\
-c_{1}s_{2}c_{3}+s_{1}c_{3} & -\left(
c_{1}s_{1}+s_{1}s_{2}c_{3}\right)  &
c_{2}c _{3}
\end{array}
\right)
\end{equation}
and $c_i=\cos\alpha_i$, $s_i=\sin\alpha_i$ for $-\frac{\pi}{2}\leq\alpha_{1,2}\leq\frac{\pi}{2}$ and $0\leq\alpha_3\leq\frac{\pi}{2}$.
The $\eta_{1,2}$ denote the real parts of the complex scalar field in weak-eigenstate, $\phi^0_{a}=\frac{1}{\sqrt{2}}\left(v_a+\eta_a+i\chi_a\right)$, whereas $\eta_3$ is written in terms of the imaginary parts and is orthogonal to the Goldstone boson, such as $\eta_3=-\chi_1\sin\beta+\chi_2\cos\beta$. The neutral Higgs bosons $h_i$ are defined to satisfy the masses hierarchy given by the inequalities $m_{h_1}\leq m_{h_2}\leq m_{h_3}$ \cite{rot,moretti}.
%
%
%%%%%%%%%%%%%%%%%%%%%%%%%%%%%%%%%%%%%%%%%%%%%%%%%%%%%%%%%%%%%%%%%%%% SECTION 3 %%%%%%%%%%%%%%%%%%%%%%%%%%%%%%%%%%%%%%%%%%%%%%%
\section{Yukawa interactions with neutral scalar-pseudoscalar mixing}
\label{sec3}
Now, we will describe the interactions between fermions and neutral Higgs bosons. The most general structure of the Yukawa interactions for fermions fields can be written as follows:
\begin{eqnarray}
-\mathcal{L}_{Yukawa} &=&\sum_{i,j=1}^{3}\sum_{a=1}^{2}\left( \overline{q}%
_{Li}^{0}Y_{aij}^{0u}\widetilde{\Phi }_{a}u_{Rj}^{0}+\overline{q}%
_{Li}^{0}Y_{aij}^{0d}\Phi _{a}d_{Rj}^{0}\right.   \nonumber \\
&&\left. +\overline{l}_{Li}^{0}Y_{aij}^{0l}\Phi _{a}e_{Rj}^{0}+h.c.\right) ,
\label{yukawa}
\end{eqnarray}
where $Y_{a}^{u,d,l}$ are the $3\times 3$ Yukawa matrices. The $q_{L}$ and $l_{L}$
denote the left handed fermions doublets meanwhile $u_{R}$, $d_{R}$, $l_{R}$ correspond to the right handed singlets. The zero superscript in fermions fields stands for weak eigenstates. After getting a correct spontaneous symmetry breaking by using (\ref{vev}), the mass matrices become
\begin{equation}
M^{u,d,l}=\sum_{a=1}^{2}\frac{v_{a}}{\sqrt{2}}Y_{a}^{u,d,l},  \label{mass}
\end{equation}
where $Y_a^{f}=V_L^f Y_a^{0f}\left(V_R^{f}\right)^\dag$ for $f=u,d,l$. The $V_{L,R}^f$ matrices are used to diagonalize the fermions mass matrices and relate the physical and weak states. If general scalar potential is considered, the neutral Higgs fields are CP-even and CP-odd mixing states as we discussed previously. In order to study the rare top decay we are interested in up-quarks and charged leptons fields. By using equations (\ref{h-Rn}), the interactions between neutral Higgs bosons and fermions can be written in the form of the 2HDM type II with additional contributions which arise from Yukawa couplings $Y_1$ and contain flavor change. In order to simplify the notation we will omit the subscript 1 in Yukawa couplings. Explicitly we write the interactions for up-type quarks and neutral Higgs bosons as
\begin{eqnarray}
\mathcal{L}_{h_{k}}^{up-quarks} &=&\frac{1}{v\sin \beta }\sum_{i,j,k}\left(
R_{k2}-i \gamma _{5}R_{k3}\cos \beta \right) \overline{u}_{i}M_{ij}^{u}h_{k}u_{j}
\nonumber \\
&&-\frac{1}{\sqrt{2}\sin \beta }\sum_{i,j,k}\left( R_{k1}\sin \beta
+R_{k2}\cos \beta \right.   \nonumber \\
&&\left. -i\gamma _{5}R_{k3}\right) \overline{u}_{i}Y_{ij}^{u}h_{k}u_{j}
\label{yukawa_quarks}
\end{eqnarray}
meanwhile the interactions for charged leptons and neutral Higgs bosons are
\begin{eqnarray}
\mathcal{L}_{h_{k}}^{leptons} &=&-\frac{1}{v\sin \beta }\sum_{i,j,k}\left(
R_{k2}+i\gamma _{5} R_{k3}\cos \beta \right) \overline{e}_{i}M_{ij}^{l}h_{k}e_{j}
\nonumber \\
&&-\frac{1}{\sqrt{2}\sin \beta }\sum_{i,j,k}\left( R_{k1}\sin \beta
-R_{k2}\cos \beta \right.   \nonumber \\
&&\left. -i\gamma _{5}R_{k3}\right) \overline{e}_{i}Y_{ij}^{l}h_{k}e_{j}.
\label{yukawa_leptons}
\end{eqnarray}
The fermion spinors are denoted as $(u_1,\,u_2,\,u_3)=(u,\,c,\,t )$ and $(e_1,\,e_2,\,e_3)=(e,\,\mu,\,\tau )$. The down-type quarks are analogous to charged leptons sector. We note that (\ref{yukawa_quarks}) and (\ref{yukawa_leptons}) generalize expressions obtained by \cite{rot,moretti,roberto1,roberto2}. The CP conserving case is obtained if only two neutral Higgs bosons are mixed with well-defined CP states, for instance for $\alpha_2=0$ and $\alpha_3=\pi/2$ is the usual limit.
%
%
%%%%%%%%%%%%%%%%%%%%%%%%%%%%%%%%%%%%%%%%%%%%%%%%%%%%%%%%%%%%%%%%%%%% SECTION 4 %%%%%%%%%%%%%%%%%%%%%%%%%%%%%%%%%%%%%%%%%%%%%%%
\section{Rare top decay through neutral Higgs bosons}
\label{sec4}
We assume that the flavor neutral changing Higgs interactions are responsible for rare top decay at tree level. The mass of the lightest physical Higgs boson $h_1$ is identified with the observed particle by ATLAS and CMS with a mass value of the order of 125 GeV, meanwhile the masses of $h_{2,3}$ are considered in region of more than 600 GeV. Then, contributions of physical neutral Higgs bosons $h_{2,3}$ are neglected in the amplitude for the width of rare top decay and only the contributions of the lightest neutral Higgs boson are taken into account. Therefore, width for rare top decay at tree level is given by
\begin{equation}
\frac{d\Gamma_{t\rightarrow c l^+l^-}}{dxdy} =\frac{m_{t}\left\vert G_{23}^{u}\right\vert
^{2}\left\vert G_{ii}^{l}\right\vert ^{2}}{128\pi ^{3}}\frac{\left( 1+\mu
_{c}-x\right) \left( x+2\sqrt{\mu _{c}}\right) }{\left( 1+\mu _{c}-\mu
_{h}-x\right) ^{2}+\mu^2 _{\Gamma}},  \label{gamma_tcll}
\end{equation}%
where
\begin{equation}
\left\vert G_{23}^{u}\right\vert ^{2}=\frac{\left\vert Y_{23}^{u}\right\vert
^{2}}{2\sin ^{2}\beta }\left[ \left( R_{11}\sin \beta -R_{12}\cos \beta
\right) ^{2}+R_{13}^{2}\right]   \label{g23}
\end{equation}
and
\begin{eqnarray}
\left\vert G_{ii}^{l}\right\vert ^{2} &=&\frac{1}{2\sin ^{2}\beta }\left[
Y_{ii}^{l}\left( R_{11}\sin \beta -R_{12}\cos \beta \right) +\sqrt{2}\frac{%
m_{i}}{v}R_{13}^{2}\right] ^{2}  \nonumber \\
&&+\frac{R_{13}^{2}}{2\sin ^{2}\beta }\left( Y_{ii}^{l}-\sqrt{2}\frac{m_{i}}{%
v}\cos \beta \right) ^{2}.  \label{g_ii}
\end{eqnarray}
In the expression for width decay (\ref{gamma_tcll}) we have used the usual notation for dimensionless parameters, 
$\mu_{c}=m^2_{c}/m^2_{t}$, $\mu_{h}=m^2_{h_1}/m^2_{t}$, $\mu_{\Gamma}=\Gamma_{H} m_{h_1}/m^2_{t}$, $x=2E_{c}/m_{t}$ and $y=2E_{l}/m_{t}$. 
We note that $m_{h_1}^2$ can be of the same order as 
the square of transferred momentum, then our result is computed without approximation in the propagator.
By integrating the expression (\ref{gamma_tcll}) we can estimate the branching ratio for $t\rightarrow c l^+l^-$. We use the experimental mean value for full width of the top quark given by $\Gamma_{t}\approx 1.6$ GeV  and the width of the Higgs field given by $\Gamma_{h_1}\approx 1.6$ GeV\cite{width-top}. Additionally, we assume that the Yukawa matrices have the structure based in Sher-Cheng ansatz \cite{2hdmIII,Fritzsch:1977za}, that is, $Y_{ij}^f\sim \sqrt{m_im_j}/v$. Therefore, the resulting branching ratio only has dependence on $\alpha_1$ and $\alpha_2$. The $\alpha_3$ mixing angle is absent in the physical state for $h_1$. Allowed regions for the $\alpha_1-\alpha_2$ parameter space are obtained through the bounds of the $R_{\gamma\gamma}$, defined by
\begin{equation}
R_{\gamma\gamma}=\frac{\sigma(gg\rightarrow h_1)Br(h_1\rightarrow\gamma\gamma)}{ \sigma(gg\rightarrow h_{SM})Br(h_{SM}\rightarrow\gamma\gamma)}.
\end{equation}
For charged Higgs boson with mass of the order of $100$ GeV - $300$ GeV, the $Br(h_1\rightarrow\gamma\gamma)$ contains an important contribution from charged Higgs boson at one level loop, which affects the allowed regions for $\alpha_1-\alpha_2$. Thus, it is possible to find allowed values in the $\alpha_1-\alpha_2$ parameter space if the parameters $\beta$ and $m_{H^\pm}$ are fixed. A process used to set $\tan\beta$ and charged Higgs boson mass is, for instance, the flavor changing process $B\rightarrow \chi_s\gamma$ \cite{misiak2012} that receives a contribution from 2HDM through charged Higgs boson. This contribution is comparable to the contribution of $W^\pm$ from SM. For small values of $\tan\beta$ this process gives a bound to the charged Higgs boson mass of the order of 300 GeV \cite{ref67,ref65}. Contributions from other processes such as $B_\tau\rightarrow\tau\nu_\tau$, $B\rightarrow D\tau\nu_\tau$, $Z\rightarrow \bar{b}b$, $B_{d,s}\rightarrow\mu^+\mu-$ and $B^0-B^0$; set bounds for the mass of $H^\pm$ and $\tan\beta$ as $m_{H^\pm}<400$ GeV and $\tan\beta\leq10$.

Therefore, allowed regions f the $\alpha_{1,2}$ parameter space are obtained by experimental and theoretical constrains in the framework of the 2HDM type II with CP violation for fixed $\tan\beta$ and $m_{H^\pm}$. For $0.5\leq R_{\gamma\gamma} \leq 2$, $m_{H^\pm}=300$ GeV and $\tan\beta=1$, the $\alpha_1$-$\alpha_2$ regions are \cite{moretti}
\begin{equation}
R_{1}=\left\{ 0.67\leq \alpha _{1}\leq 0.8\right.\,\textrm{ and}\,\left. 0\leq \alpha
_{2}\leq 0.23\right\}
\end{equation}
and
\begin{equation}
R_{2}=\left\{ 0.8\leq \alpha _{1}\leq 1.14\right. \,\textrm{and}\,\left. -0.25\leq \alpha
_{2}\leq 0\right\}.
\end{equation}
%
%%%%%%%%%%%%%%%%%%%%%%%%
%
%
\begin{figure} % figuur 1
\begin{minipage}{\columnwidth}
\centering
\includegraphics[scale=0.6]{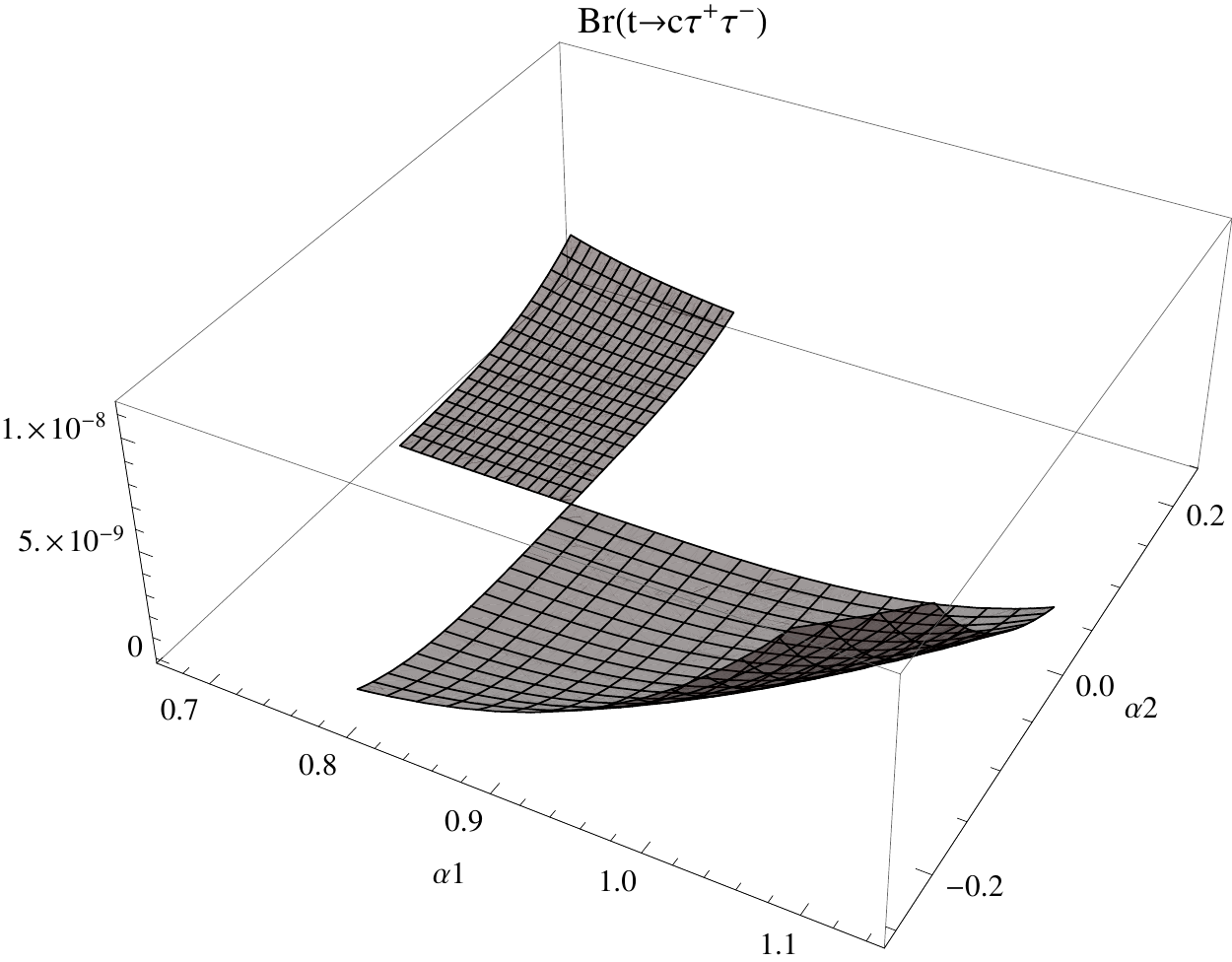}
\end{minipage}
\caption{Type III 2HDM branching ratio for $t\rightarrow c \tau^+\tau^-$ as a function of $\alpha_1$-$\alpha_2$ in regions $R_5$ and $R_6$ with $\tan\beta=1$ and $m_{H^\pm}=300$ GeV.}
\label{figure1}
\end{figure}
%
%%%%%%%%%%%%%%%%%%%%%%%%%
%
%
\begin{figure} % figuur 1
\begin{minipage}{\columnwidth}
\centering
\includegraphics[scale=0.6]{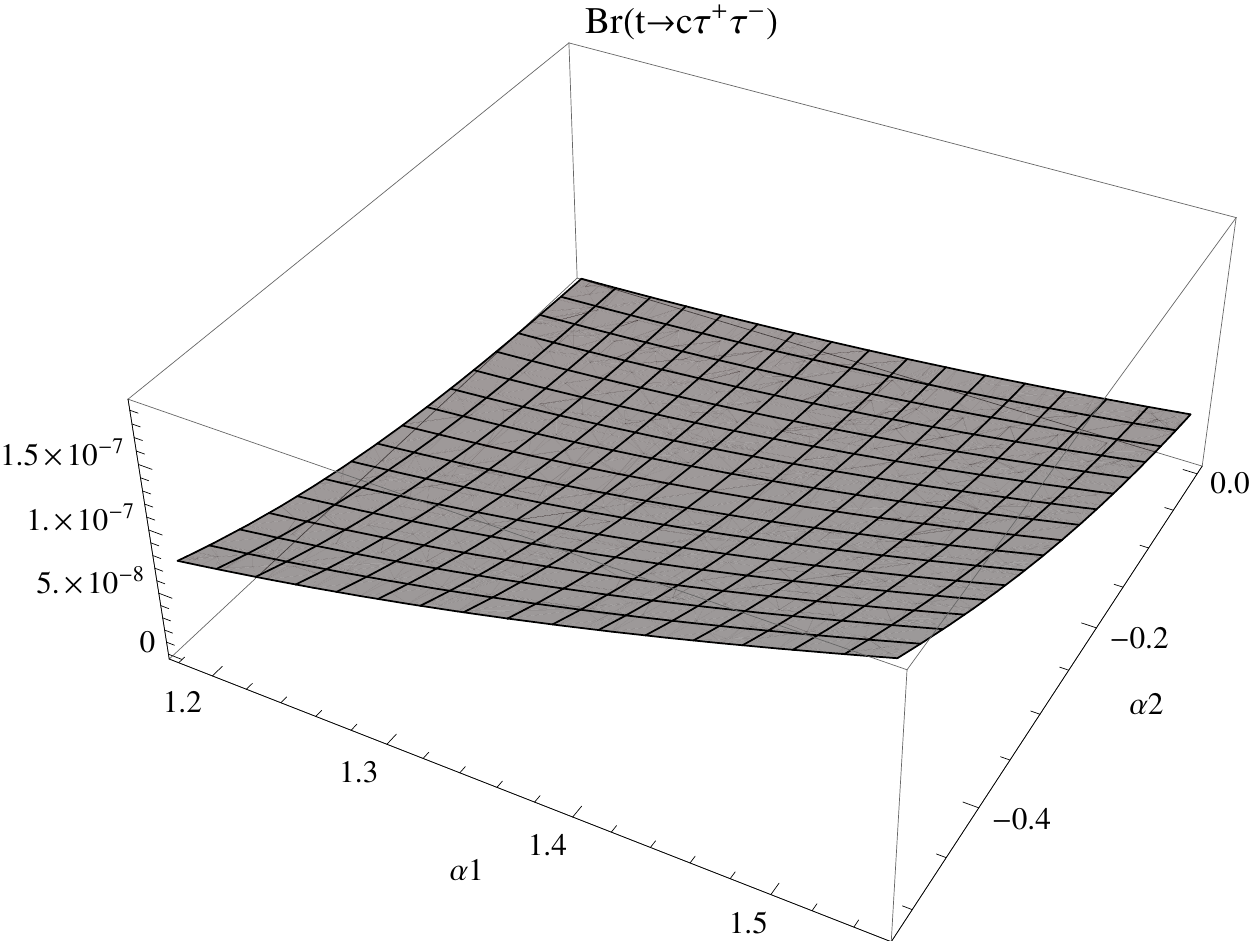}
\end{minipage}
\caption{Type III 2HDM branching ratio for $t\rightarrow c \tau^+\tau^-$ as a function of $\alpha_1$-$\alpha_2$ in region $R_3$ with $\tan\beta=1$ and $m_{H^\pm}=500$ GeV..}
\label{figure2}
\end{figure}
For the same settings but with $m_{H^\pm}=500$ GeV,
\begin{equation}
R_{3}=\left\{ 1.18\leq \alpha _{1}\leq 1.55\right. \,\textrm{and}\,\left. -0.51\leq \alpha
_{2}\leq 0\right\}.
\end{equation}
In order to reduce $\alpha_1$-$\alpha_2$ parameter space we consider these regions as an approximation. In addition, we will consider $\tau^+$ and $\tau^-$ in the final state. The Figure \ref{figure1} shows the branching ratio of rare top decay for regions $R_1$ and $R_2$ meanwhile figure \ref{figure2} is obtained for $R_3$. For $1\leq R_{\gamma\gamma} \leq 2$, $m_{H^\pm}=350$ GeV and $\tan\beta=1.5$ the allowed parameter regions in $\alpha_1$-$\alpha_2$ plane in the framework of 2HDM with potential but softly broken $Z_2$ discrete symmetry are \cite{Arhrib2012}
\begin{equation}
R_{4}=\left\{ -1.57\leq \alpha _{1}\leq -1.3\right. \,\textrm{and}\,\left. -0.46\leq
\alpha _{2}\leq 0\right\}
\end{equation}
and
\begin{equation}
R_{5}=\left\{ 0.93\leq \alpha _{1}\leq 1.57\,\,\right. \textrm{and} \left. -0.61\leq
\alpha _{2}\leq 0\right\}.
\end{equation}
For $\tan\beta=2$ the regions are
\begin{equation}
R_{6}=\left\{ -1.57\leq \alpha _{1}\leq -1.28\right. \textrm{and} \left. -0.38\leq
\alpha _{2}\leq 0\right\}.
\end{equation}
and
\begin{equation}
R_{7}=\left\{1.08\leq \alpha _{1}\leq 1.57\,\,\right. \textrm{and} \left. -0.46\leq
\alpha _{2}\leq 0\right\}.
\end{equation}
Finally, for $\tan\beta=2.5$ the region is
\begin{equation}
R_{8}=\left\{ -1.39\leq \alpha _{1}\leq -1.3\right. \textrm{and} \left. -0.13\leq
\alpha _{2}\leq 0\right\}.
\end{equation}
and
\begin{equation}
R_{9}=\left\{ 1.16\leq \alpha _{1}\leq 1.5\,\,\right. \textrm{and} \left. -0.43\leq
\alpha _{2}\leq -0.1\right\}.
\end{equation}
The figures \ref{figure3}, \ref{figure4} and \ref{figure5} show the branching ratio for previous regions.
We note that the branching ratio of rare top decay for $\tan\beta=1$ and $m_{H^\pm}=500$ GeV is bounded as $Br(t\rightarrow c \tau^+\tau^-)\leq 5\times 10^{-7}$ for any $\alpha_{1,2}$. For $\mu^+$ and $\mu^-$ pair in final state we find that $Br(t\rightarrow c \mu^+\mu^-)\leq 1.9\times 10^{-9}$ with same $\tan\beta=1$. If $\beta$ mixing angle is fixed with values greater than $\tan\beta=1$, the branching ratio does not vary drastically over all $\alpha_1$-$\alpha_2$ region; for instance when $\tan\beta=45$ then $Br(t\rightarrow c \tau^+\tau^-)\leq 2.8\times 10^{-7}$. The table \ref{tabla1} contains the upper bounds for the considered regions.
\begin{table}
\centering
\caption{Maximum numerical value of the $Br(t\rightarrow c l^+l^-)$ for the considered regions. Last column contains a naive estimation for the events that could be observed with luminosity of the order of 300 $\textrm{fb}^{-1}$ and 14 GeV for the center of mass energy.}
\label{tabla1}
\begin{tabular*}{\columnwidth}{@{\extracolsep{\fill}}lll@{}}
\hline
Regions & Upper bound  & Events\\
\hline
$R_1$ & $2.52 \times 10^{-9}$ & 0 \\
\hline
$R_2$ & $1.24 \times 10^{-8}$ & 0 \\
\hline
$R_3$ & $1.93 \times 10^{-7}$ &  10\\
\hline
$R_4$ & $3.22 \times 10^{-8}$ &  2\\
\hline
$R_5$ & $8.46 \times 10^{-8}$ &  4\\
\hline
$R_6$ & $1.61 \times 10^{-8}$ &  1\\
\hline
$R_7$ & $2.84 \times 10^{-8}$ &  1\\
\hline
$R_8$ & $8.55 \times 10^{-9}$ &  0\\
\hline
$R_9$ & $1.66 \times 10^{-8}$ &  1\\
\hline
\end{tabular*}
\end{table}
%
%%%%%%%%%%%%%%%%%%%%%%%%
%
%
\begin{figure} % figuur 1
\begin{minipage}{\columnwidth}
\centering
\includegraphics[scale=0.6]{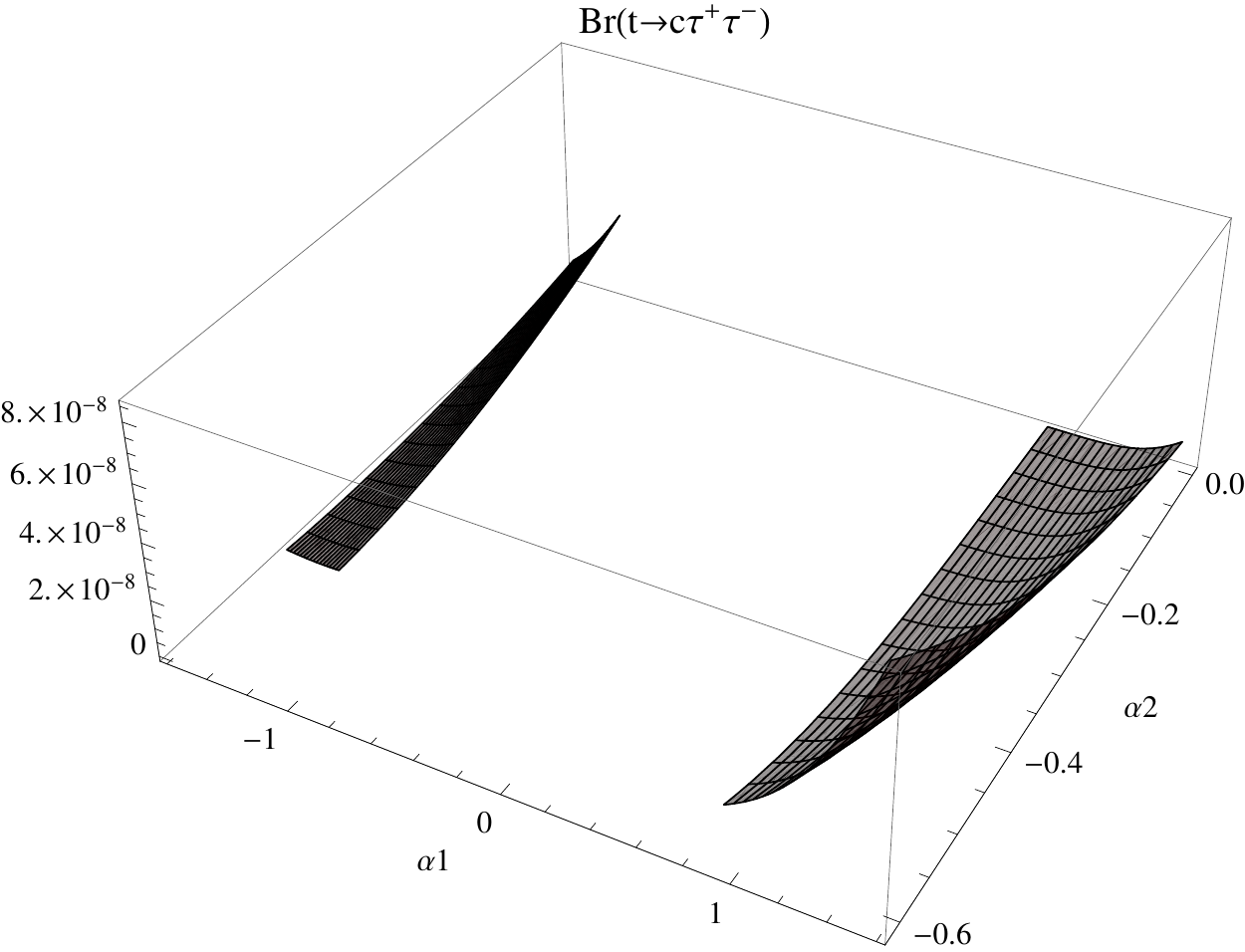}
\end{minipage}
\caption{Type III 2HDM branching ratio for $t\rightarrow c \tau^+\tau^-$ as a function of $\alpha_1$-$\alpha_2$ in regions $R_4$ and $R_5$  with $\tan\beta=1.5$ and $m_{H^\pm}=350$ GeV..}
\label{figure3}
\end{figure}
%
%%%%%%%%%%%%%%%%%%%%%%%%%
%
%
\begin{figure} % figuur 1
\begin{minipage}{\columnwidth}
\centering
\includegraphics[scale=0.6]{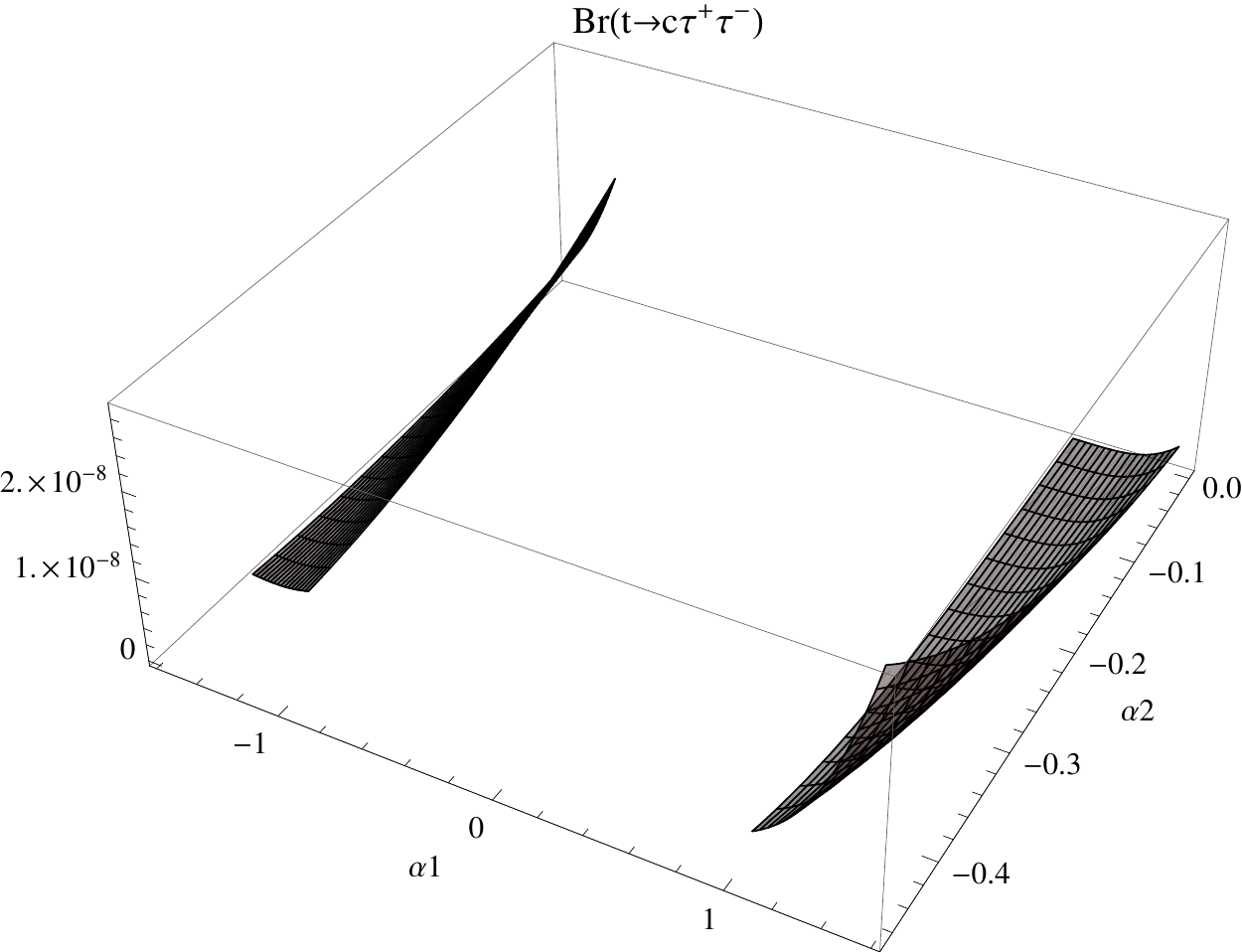}
\end{minipage}
\caption{Type III 2HDM branching ratio for $t\rightarrow c \tau^+\tau^-$ as a function of $\alpha_1$-$\alpha_2$ in regions $R_6$ and $R_7$  with $\tan\beta=2$ and $m_{H^\pm}=350$ GeV..}
\label{figure4}
\end{figure}
%
%
%
%%%%%%%%%%%%%%%%%%%%
%%%%%%%%%%%%%%%%%%%%%%%%
%
%
\begin{figure} % figuur 1
\begin{minipage}{\columnwidth}
\centering
\includegraphics[scale=0.6]{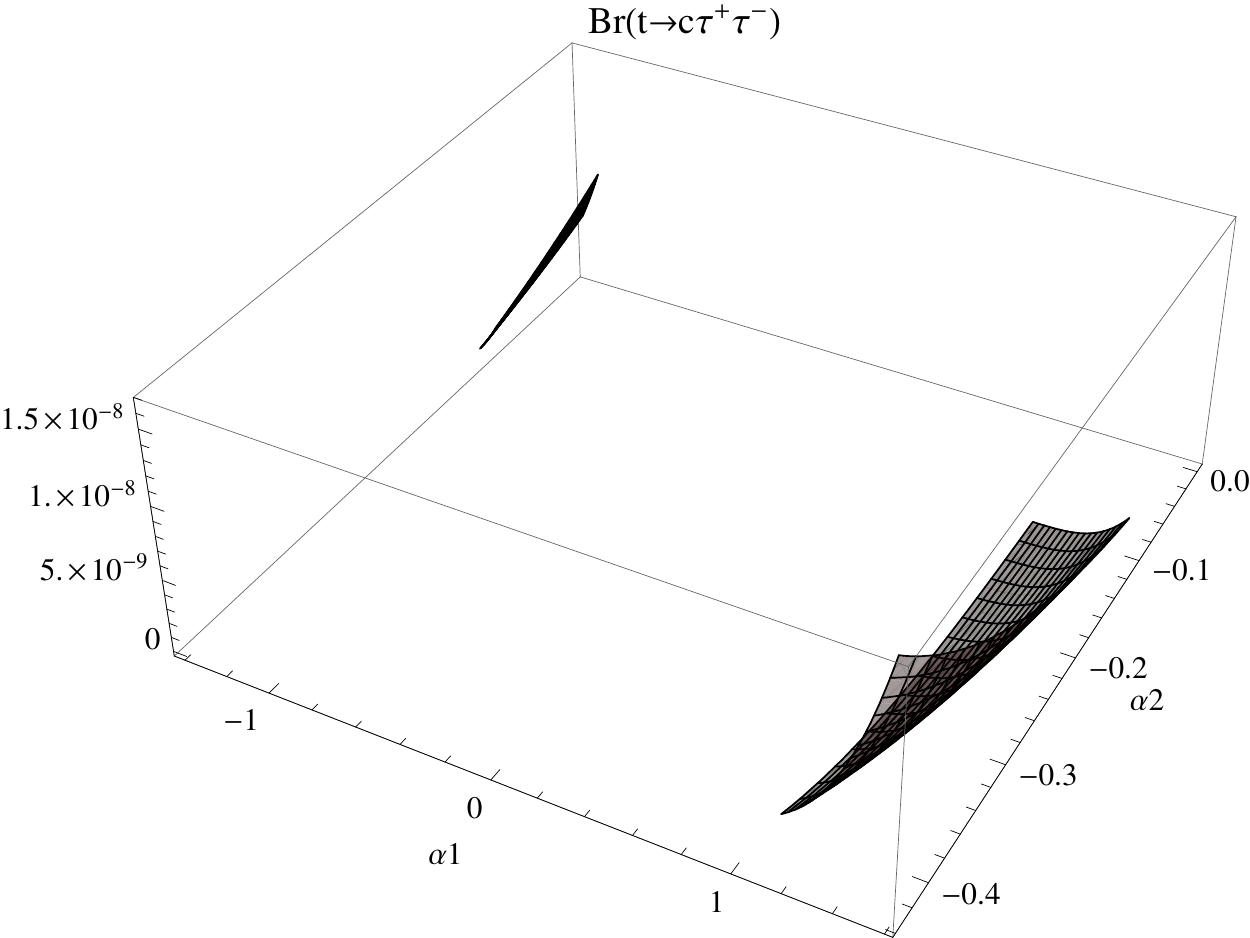}
\end{minipage}
\caption{Type III 2HDM branching ratio for $t\rightarrow c \tau^+\tau^-$ as a function of $\alpha_1$-$\alpha_2$ in regions $R_8$ and $R_9$  with $\tan\beta=2.5$ and $m_{H^\pm}=350$ GeV..}
\label{figure5}
\end{figure}
%
%
%%%%%%%%%%%%%%%%%%%%%%%%%%%%%%%%%%%%%%%%%%%%%%%%%%%%%%%%%%%%%%%%%%%% SECTION 4 %%%%%%%%%%%%%%%%%%%%%%%%%%%%%%%%%%%%%%%%%%%%%%%
\section{Discussion and conclusion}
\label{sec5}
From 2015 to 2017 the experiment is expected to reach 100 $\textrm{fb}^{-1}$ of data with a energy of the center of mass of 14 TeV. In the year 2021 is expected to reach a luminosity of the order of 300 $\textrm{fb}^{-1}$ of data. Experiments with this luminosity could find evidence of new physics beyond SM. Then, Run 3 in LHC could observe events for the neutral flavor changing process such that $t\rightarrow c h \rightarrow c l^+l^-$, which can be explained in a naive form as
\begin{eqnarray}
Br(p\bar{p} &\rightarrow &\bar{b}Wcl^{+}l^{-})  \nonumber \\
&\approx &\sigma (p\bar{p}\rightarrow t\bar{t})Br(\bar{t}\rightarrow \bar{b}%
W)Br(t\rightarrow cl^{+}l^{-}).
\end{eqnarray}
Then, we estimate the number of events using the upper bound for branching ratio with $\sigma(p\bar{p}\rightarrow t\bar{t})\approx 176 \,pb $ \cite{pdg2012}. The table \ref{tabla1} contains this estimation for the considered regions.

Finally, we compare our result with reported results in others framework, such as effective theories and 2HDM type I or II. Based on (\ref{yukawa_quarks}) we can write the branching ratio for $t\rightarrow c h_1$ as
\begin{eqnarray}
Br(t &\rightarrow &ch_{1})  \nonumber \\
&=&\frac{m_{t}\left\vert G_{23}^{u}\right\vert ^{2}}{4\pi \Gamma _{t}}\sqrt{%
\lambda \left( 1,\mu _{c},\mu _{h}\right) }\left( 1-\mu _{c}-\mu _{h}-\sqrt{%
\mu _{c}}\right)   \label{br-tch}
\end{eqnarray}
where $\lambda$ is the usual function. We find that $Br(t\rightarrow ch_{1})\leq 5 \times 10 ^{-3}$ with $m_{h_1}=125$ GeV and $\tan\beta=1$.
%
%\begin{figure} % figuur 1
%\begin{minipage}{\columnwidth}
%\centering
%\includegraphics[scale=0.6]{br_tch.pdf}
%\end{minipage}
%\caption{Branching ratio for rare top decay $t\rightarrow c h_1$ for $m_{h_1}=125$ GeV and $\tan\beta=1$.}
%\label{fig-br-tch}
%\end{figure}
%
%
%\begin{figure} % figuur 1
%\begin{minipage}{\columnwidth}
%\centering
%\includegraphics[scale=0.6]{h1tau.pdf}
%\end{minipage}
%\caption{Branching ratio for $h_1 \rightarrow \tau^+\tau^-$ for $m_{h_1}=125$ GeV and $\tan\beta=1$.}
%\label{fig-br-h1tau}
%\end{figure}
%
Despite the absence of flavor changing neutral Higgs interactions in SM, $t\rightarrow c h_{SM}$ decay can occur at one loop level. The reported result for the branching ratio is of the order of $10^{-14}$-$10^{-13}$ for $m_{Z}\leq m_{SM} \leq 2m_{Z}$ \cite{mele}. More recently, in the framework of general 2HDM with CP-even $(H^0)$ and CP-odd $(A^0)$ neutral Higgs bosons the branching ratios are estimated as $\textrm{Br}(t\rightarrow c H^0)=2.2\times 10 ^{-3}$ and $\textrm{Br}(t\rightarrow c A^0)=1.2\times 10 ^{-4}$ for $m_{H^0}=125$ GeV and $m_{A^0}=150$ GeV \cite{kao2012}. By using effective operator formalism the flavor changing neutral Higgs interactions are introduced. An upper bound is estimated as $\textrm{Br}(t\rightarrow c H^0)=2.7 \% $ for neutral Higgs mass of 125 GeV \cite{Craig2012}. Top decays with effective theories is also studied, for the case of $t\rightarrow c h$ the $\textrm{Br}(t\rightarrow c H^0)=5\times 10 ^{-3}$ for $m_h=125$ GeV are obtained \cite{Toscano2010}. In  reference \cite{Roberto2005} has been estimated upper bound $\textrm{Br}(t\rightarrow c H)=0.09-2.8\times 10 ^{-3}$ for $114\leq m_H\leq 170$ GeV through the one loop contributions of effective flavor changing neutral couplings $tcH$ on the electroweak precision observables in SM. For Yukawa complex couplings and CP effects in 2HDM type III the $\textrm{Br}(t\rightarrow c H^0)\approx 10 ^{-3}$ is predicted by \cite{iltan}. 

From reference \cite{1205} fugure 3 can be estimated the branching ratio of the $h_1$ into $\tau$'s which is the order of $BR(h_1\to \tau\tau)\approx 0.05$ for any value of $\alpha_1$ and $\alpha_2$. Using this BR and taking into account $BR(t\to ch_1)\leq 10^{-3}$ for different scenarios of models, we obtain 
\begin{equation}
BR(t\to ch_1\to tc\tau\tau)\approx 5\times10^{-5}
\end{equation}
which is two order of magnitude bigger than the value obtain for us for different regions of parameters, table 1. The number of events, in the best scenario, at LHC with $300 fb^{-1}$ of luminosity and $14$ TeV for the energy of the center of mass is the order of $2500$.

%
%%%%%%%%%%%%%%%%%%%%%%%%%%%%%%%%%%%%%%%%%%%%%%%%%%%%%%%%%%%%%%%%%%% A Q U I
%
%
%
%%%%%%%%%%%%%%%%%%%%%%%%%%%%%%%%%%%%%%%%%%%%%%%%%%%%%%%%%%%%%%%%%%%% SECTION Acknowledgments %%%%%%%%%%%%%%%%%%%%%%%%%%%%%%%%%%%%%%%%%%%%%%%
%
\section*{Acknowledgments}
This work is supported in part by PAPIIT project IN117611-3, Sistema
Nacional de Investigadores (SNI) in M\'exico. J.H. Montes de Oca Y.
is thankful for support from the postdoctoral DGAPA-UNAM grant. R. M. thanks to COLCIENCIAS for the financial support.
%
%
%\appendix
%\section{The integrals $I_i$}
%
%
%%%%%%%%%%%%%%%%%%%%%%%%%%%%%%%%%%%%%%%%%%%%%%%%%%%%%%%%%%%%%%%% R E F E R E N C E %%%%%%%%%%%%%%%%%%%%%%%%%%%%%%%%%%%%%%%%%%%%%%%%%%%%%%%555
%\section*{References}

\end{document}